\documentclass[preprint,5p,times,twocolumn]{elsarticle}
\usepackage{graphicx,epstopdf}
\usepackage{multicol,setspace}
\usepackage{float}
\usepackage{hyperref}
\usepackage{color,xcolor}
\usepackage{epsfig}
\usepackage{caption}
\usepackage{amsmath,amssymb,amsfonts}
\usepackage[titletoc,toc,page]{appendix}
\usepackage{lipsum}
\usepackage{lineno}
\def\beq{\begin{equation}}
\def\eeq{\end{equation}}
\def\beqn{\begin{eqnarray}}
\def\eeqn{\end{eqnarray}}
\newcounter{saveeqn}

\def\r {{\bf r}}

\def\C {{\bf C}}

\def\brho {\mbox{\boldmath $\rho$}}

\journal{Chemical Physics}

\begin{document}

\begin{frontmatter}
\title{Impact of the range of the interaction on the quantum dynamics of a bosonic Josephson junction}
\author[1,2]{Sudip Kumar Haldar\corref{cor1}}
\ead{shaldar@campus.haifa.ac.il}
\cortext[cor1]{Corresponding author.}
\author[1,2]{Ofir E. Alon}
\ead{ofir@research.haifa.ac.il}
\address[1]{Department of Mathematics, University of Haifa, Haifa 3498838, Israel}
\address[2]{Haifa Research Center for Theoretical Physics and Astrophysics,
University of Haifa, Haifa 3498838, Israel}


\begin{abstract}
The out-of-equilibrium quantum dynamics of a bosonic Josephson junction (BJJ) with long-range interaction is studied in real space by solving the time-dependent 
many-body 
Schrödinger equation numerically accurately using the multiconfigurational time-dependent Hartree method for bosons. Having the many-boson wave-function at 
hand we can examine 
the impact of the range of the interaction on the properties of the BJJ dynamics, viz. density oscillations and their collapse, self trapping, 
depletion and fragmentation, as well as 
the position variance, both at the mean-field and many-body level. 
Explicitly, the frequency of the density oscillations and the time required for 
their collapse, the value of fragmentation at the plateau, the maximal and the minimal values of the position variance 
in each cycle of oscillation and the overall pace of its growth are key to our study.
We find competitive effect between the interaction and the confining trap. The presence of 
the tail part of the interaction basically enhances the effective repulsion as the range of the interaction is increased starting from a short, finite range. 
But, as the range becomes comparable with the trap size, the system approaches a situation where all the atoms feel a constant potential  and 
the impact of the tail on the 
dynamics diminishes. There is an optimal range of the interaction in which physical quantities of the junction are attaining their extreme values.
\end{abstract}



\end{frontmatter}

\section{Introduction}
\label{Introduction}
The recent advancement{s} in experimental techniques for interacting Bose gas {have} made it possible to study the 
quantum many-body dynamics in a highly controllable manner~\cite{Dalfovo}. 
In this connection, the dynamics of many-body tunneling~\cite{Meyer2008} is one of the most fundamental problem. 
{A} Bose-Einstein condensate (BEC) of interacting dilute Bose gas in a double well, 
which is generally referred to as {a} bosonic Josephson junction (BJJ)~\cite{Gati}, provides a unique 
opportunity to study many-body tunneling dynamics. Naturally, the dynamics 
of {BJJs have} attracted a lot of attention both theoretically and 
experimentally~\cite{Gati, Milburn, Shenoy, Levi, Raghavan, Ostrovskaya, Zhou, Lee, Ananikian, Gati2007, Ferrini, 
Shchesnovich, Jia, Trujillo, Sakmann2009, Sakmann2010, Zibold, Sakmann2014, Klaiman2016, Spagnolli, Burchinati, Hou}. 
Explicitly, Josephson {oscillations}~\cite{Gati, Shenoy, Levi, Zibold, Spagnolli, Burchinati}, collapse and revival 
{cycles}~\cite{Milburn}, {self trapping} (suppression of tunneling)~\cite{Gati, Milburn, Shenoy, Levi, Zibold}, 
fragmentation~\cite{Sakmann2014} and {more recently the} variances
{and uncertainty product of the} many-body position and momentum operators~\cite{Klaiman2016} have been {studied}. 
Note that while tunneling, {self trapping} and Josephson 
{oscillations} have some explanations {at the} mean-field level, {the} collapse and revival and fragmentation 
dynamics require many-body treatments like {the}
Bose-Hubbard model~\cite{Gati2007} or even solving {the} full many-body Schr\"odinger equation~{\cite{Sakmann2009, MCHB}}. 
A universality has been predicted in the {fragmentation dynamics 
in the sense that} systems consisting of different numbers of particles fragment to the same value for {the same} 
mean-field interaction {parameter}~\cite{Sakmann2014}. 
However, it has been shown that even when {the} Bose-Hubbard model is {apparently} applicable, {the} 
full many-body Schr\"odinger equation can {grab new} features. 
For example, there is a symmetry in the Bose-Hubbard 
Hamiltonian with respect to repulsive and attractive interactions of equal magnitude. Such symmetry implies an 
equivalence between the time evolution of {the} survival probability 
and the fragmentation of a repulsive and an attractive BJJ with equal magnitude of the strength of the interaction. 
However, no such symmetry exists {at the level} of {the} full many-body Hamiltonian~\cite{Sakmann2010}. 

So far, only contact interactions between the atoms have been considered {in the study of 
BJJs}~{\cite{Sakmann2009, Sakmann2010, Sakmann2014, Spagnolli, Burchinati}}. {Actually},
{contact $\delta$-interaction is widely used in the theoretical studies of trapped ultra-cold atomic gases~\cite{Dalfovo}}.
However, in many recent experiments with {the} ultra-cold {diploar atoms} $^{52}$Cr~\cite{Griesmaier, Stuhler}, 
$^{164}$Dy~\cite{Lu} and $^{168}$Er~\cite{Aikawa}, 
it has been shown that the short-range inter-particle interaction potential is not enough to {account for} the observed physics 
and an additional long-range term is {needed to {describe} the 
overall two-body interaction, {see also the reviews}~\cite{Baranov, Lahaye}.  
It is also possible, in experiments, to tune the strength of the dipolar 
interactions including its sign by using a rotating polarizing field~\cite{Giovanazzi}. 
For a $^{52}$Cr BEC, one can also use the Feshbach resonance to tune the $s$-wave scattering length~\cite{Werner}{, 
and this} has already been used to enhance the dipolar effects in {a} BEC~\cite{Lahaye2007}. }
Naturally, the question arises what role {the range of the interaction} plays and how that affects our present 
understanding of the physics of {an} {ultra-cold Bose gas}. 
 
{To address these questions, several static properties including the fragmentation of the ground state of the trapped 
ultra-cold bosonic atoms with long-range interaction  
have already been studied theoretically in one, two and three {spatial} dimensions~\cite{Bader2009, Streltsov2013, Fischer2015}. 
Also, the non-equilibrium dynamics of trapped bosons interacting by a long-range interaction has been studied both at the 
mean-field level~\cite{Baranov, Lahaye, Henkel2012, Santos} as well as the many-body level~\cite{Oksana2014}. However, 
all these studies considered mainly a single trap of various {geometries}.}

Therefore, in this work, {we would like to bring together the topics of BJJs and long-range interactions.} 
We numerically study the tunnelling dynamics 
of a BEC with a tunable long-range interactions in a double well. Here we are interested in the generic 
behaviour of the quantum many-body dynamics with a finite-range inter-atomic
interaction. The physics of the many-body dynamics is not expected to vary {qualitatively} 
with the shape of the interactions provided the strength and range of the interaction remain {the} same.
In this work we consider a model interaction $W(r)=\frac{\lambda_0}{\sqrt{(r/D)^{2n}+1}}$ of strength 
$\lambda_0$, half-width $D$ with $n=3$ and $r$ being the 
inter-particle separation. {In our study, we will vary $D$ for fixed values of $\lambda_0$.} 
Such interactions appear naturally in the so called ``Rydberg-dressed'' 
ultra-cold systems which are studied in recent experiments~\cite{Johnson, Henkel}. 
Moreover, at large $r$, $r/D \gg 1$, such interactions behave as a dipolar interaction which is relevant 
{for dipolar} ultra-cold atoms~\cite{Baranov, Lahaye}. 
On the other hand, {at small} $r$ $(r/D \ll 1)$, {$W(r)$} reduces
to a soft sphere interaction and {has} many similar effects as the usual $\delta$-interaction. 

{In our studies below, starting} with a short range, we tune the {effective} range of the interaction. 
We observe that the presence of the long-range tail in the inter-particle interaction potential 
basically enhances the effect of the interaction {until the range of the interaction becomes comparable 
with the {inter-well separation}}. Also, for the stronger 
interaction, the range of the interaction plays {a} more prominent role. Already {at the} mean-field level, 
we observe {clear} effect on the Josephson oscillation 
frequency and amplitude for a sufficiently strong interaction strength. Naturally, the many-body dynamics is even richer. 
{The loss} of coherence and development of {correlations} and 
fragmentation {are} significantly affected due to the presence of the long-range tail in the inter-particle interaction 
potential which in turn alters the usual collapse 
{of oscillations of the survival probability} 
and {self trapping}. Variances {of operators} which, unlike other quantities, depend on the actual 
number of depleted atoms and not on the condensate fraction {are} a very sensitive measure of 
{correlations}. In the {infinite-particle} limit, as the depletion tend{s} to zero and 
the energy per particle and density per particle of a condensate overlaps 
with those obtained from a mean-field theory, variances {of operators} {can} show significantly 
different behaviour which can only be obtained with a many-body theory. Naturally, in this case 
also, {the} variance reveals more prominent effect of the long-range behavior.

This paper is organised as follows. In Sec.~\ref{Theory}, we introduce {{the in principle}
numerically-exact many-body method} {used to solve the time-dependent} many-body Schr\"odinger equation. In 
sec.~\ref{Result} we discuss our findings and finally {conclusions are} drawn in Sec.~\ref{conclusion}. 
{The Appendix puts forward additional numerical details}.

\section{Theoretical framework}
\label{Theory}

The time evolution of $N$ interacting structureless bosons is governed by the time-dependent many-body Schr\"odinger equation:
\beq \label{MBSE}
 \hat H \Psi = i \frac{\partial \Psi}{\partial t}, \qquad 
 \hat H(\r_1,\r_2,\ldots,\r_N) =  \sum_{j=1}^{N} \hat h(\r_j) +  \sum_{k>j=1}^N W(\r_j-\r_k),
 \eeq
where $\hbar=1$, $\r_j$ {is} the coordinate of the $j$-th boson, $\hat h(\r) = \hat T(\r) + V(\r)$ is the one-body Hamiltonian 
containing kinetic and potential energy terms, {and}
$W(\r_j-\r_k)$ is the pairwise interaction between the $j$-th and $k$-th bosons.
The time-dependent many-boson Schr\"odinger equation (\ref{MBSE}) cannot be solved exactly (analytically), 
except for a few specific cases only, see, e.g.,~\cite{Marvin}.
Hence, {an in principle numerically-exact} many-body method {for identical bosons, based on the multi-configurational 
time-dependent Hartree (MCTDH)~\cite{Meyer1990,Meyer1992} method, 
viz. the multi-configurational time-dependent Hartree method for bosons (MCTDHB)}{,} was developed to solve Eq.~(\ref{MBSE}).  
Detailed derivation of the MCTDHB equation of motions can be found in~\cite{Streltsov2007, Ofir2008}. 
Below we briefly describe the basic idea behind the method.

In MCTDHB, the ansatz for solving Eq.~(\ref{MBSE}) is taken as the superposition of all 
$\begin{pmatrix} 
N+M-1\\
N
\end{pmatrix}$
time-dependent permanents obtained by distributing $N$ bosons in $M$ single{-}particle 
orbitals. Thus, the many-body wavefunction $\Psi(t)$ is given by
  \beq\label{MCTDHB_Psi}
\left|\Psi(t)\right> = 
\sum_{\vec{n}}C_{\vec{n}}(t)\left|\vec{n};t\right>,
\eeq
where the summation runs over all possible configurations whose occupations $\vec{n}$ preserve the total number of bosons $N$. 
{The} time-dependent permanents are 
given by
\beq\label{basic_permanents}
 \left|\vec{n};t\right> =
\frac{1}{\sqrt{n_1!n_2!\cdots n_M!}} \left(b_1^\dag(t)\right)^{n_1}\left(b_2^\dag(t)\right)^{n_2}
\cdots\left(b_M^\dag(t)\right)^{n_M}\left|vac\right>,
\eeq
where $\vec{n}=(n_1,n_2,\cdots,n_M)$ represents the occupations of the orbitals
{such that $n_1+n_2+\cdots+n_M=N$}, 
and $\left|vac\right>$ is the vacuum. The bosonic annihilation and corresponding creation operators 
obey the usual commutation relations 
$b_k(t) b^\dag_j(t) - b^\dag_j(t) b_k(t) = \delta_{kj}$
at any {point in} time. 
Note that in representation (\ref{MCTDHB_Psi})
both the expansion coefficients $\{C_{\vec{n}}(t)\}$ and 
orbitals $\{\phi_k(\r,t)\}$ comprising the permanents $\left|\vec{n};t\right>$
are independent parameters. Throughout this work we have performed all computations with $M=2$ orbitals. 
{The results are found to be accurate for the quantities and propagation {times} considered here.}
For further details on numerical computations and its accuracy, we refer {the reader to} {the} 
Appendix.

To solve for the time-dependent wavefunction $\Psi(t)$  
we need to determine the evolution of the coefficients $\{C_{\vec{n}}(t)\}$ 
and orbitals $\{\phi_k(\r,t)\}$ with time.
To derive the equations of motion governing the evolution 
of $\{C_{\vec{n}}(t)\}$ and $\{\phi_k(\r,t)\}$, we employ the usual Lagrangian formulation of the time-dependent 
variational principle \cite{LF1,LF2} subject to the orthonormality 
between the orbitals. In this framework,
we substitute the many-body {\it ansatz} (\ref{MCTDHB_Psi}) for $\Psi(t)$
into the functional action of the time-dependent Schr\"odinger equation which reads:
\beqn\label{action_functional}
 S\left[\{C_{\vec{n}}(t)\},\{\phi_k(\r,t)\}\right] = \nonumber \\
 \int dt \left\{ \left<\Psi\left|\hat H - i\frac{\partial}{\partial t} \right|\Psi\right>
 - \sum_{k,j}^{M} \mu_{kj}(t)\left[\left<\phi_k|\phi_j\right>
 - \delta_{kj} \right]\right\}. 
\eeqn
The time-dependent Lagrange multipliers $\mu_{kj}(t)$ ensure that
the time-dependent orbitals $\phi_k(\r,t)$ remain orthonormal throughout the propagation~\cite{Ofir2008}.
The next step is to require stationarity of the {functional}
action with respect to its arguments $\{C_{\vec{n}}(t)\}$ and $\{\phi_k(\r,t)\}$.
This variation is performed separately for
the coefficients and for the orbitals,
recalling that they are independent variational parameters. After a lengthy but straightforward calculation we arrive 
at the final equations of motion which read
for the orbitals $\phi_j(\r,t)$, $j=1,\ldots,M$:
\beqn\label{MCTDHB1_equ}
& & 
  i\left|\dot\phi_j\right> = \hat {\mathbf P} \left[\hat h \left|\phi_j\right>  + \sum^M_{k,s,q,l=1} 
  \left\{\brho(t)\right\}^{-1}_{jk} \rho_{ksql} \hat{W}_{sl} \left|\phi_q\right> \right], \nonumber \\ 
& &  
\qquad \hat {\mathbf P} = 1-\sum_{j^{\prime}=1}^{M}\left|\phi_{j^{\prime}}\left>\right<\phi_{j^{\prime}}\right|. \
\eeqn
Here $\brho(t)$ is the reduced one-body density matrix and 
$\rho_{ksql}=\left<\Psi\left|b_k^\dag b_s^\dag b_q b_l\right|\Psi\right>$ are the elements of 
the reduced two-body density matrix. Given the (normalized) wavefunction $\Psi(t)$, the reduced one-body density matrix is given by
\beqn\label{1RDM}
\rho(\r_1|\r_1^{\prime};t) & = &
N \int d\r_2 \ldots d\r_N \, \Psi^\ast(\r_1^{\prime},\r_2,\ldots,\r_N;t)  \nonumber \\
& \, & \times \Psi(\r_1,\r_2,\ldots,\r_N;t) \nonumber \\
& = &\sum_{j=1}^{M} n_j(t) \, \phi^{\ast{NO}}_j(\r_1^{\prime},t)\phi^{NO}_j(\r_1,t).
\eeqn
The quantities $\phi^{NO}_j(\r_1,t)$ are the natural orbitals and $n_j(t)$ the natural occupation numbers
which are time-dependent and used to define the {(time varying)} degree of condensation 
in a system of interacting bosons \cite{PeO56}. The natural occupations satisfy $\sum_{j=1}^{M} n_j = N$ 
and if only one {macroscopic} eigenvalue 
$n_1(t) \approx {\mathcal O}(N)$ exists, the system is condensed \cite{PeO56} whereas if 
there are more than one {macroscopic} eigenvalues, the BEC
is said to be fragmented {\cite{MCHB,NoS82,Spekkens99,Ueda,Sak08}}. {The microscopic occupation}{s} 
{in the higher orbitals {$f=\sum_{j=2}^{M} \frac{n_j}{N}$} 
for a condensate is known as the depletion {per particle}. On the other hand, the macroscopic} 
{occupation} {of a higher natural orbital, 
viz. $\frac{n_{j>1}}{N}$ where $n_j \approx {\mathcal O}(N)$, is called fragmentation. Thus, it
is quite obvious that for $M=2$ {orbitals}, both the depletion and fragmentation basically have the
same mathematical expression (but different physical interpretations) and, hence, {for the ease of our discussion, below 
we refer to} both of them by $f$.}  
The density of the system is the diagonal of the reduced one-body density matrix,
$\rho(\r;t) = \rho(\r|\r;t)$.

The {equations of motion} for the propagation of the coefficients is given by
\beq\label{variation2_C_matrix}
 {\mathbf H}(t)\C(t) = i\frac{\partial \C(t)}{\partial t},
\eeq   
where ${\mathbf H}(t)$ is the Hamiltonian matrix {with the elements}
$H_{\vec{n}\vec{n}'}(t) = \left<\vec{n};t\left|\hat H\right|\vec{n}';t\right>$.
{Recently} a parallel version of MCTDHB has been
implemented using a novel mapping technique~\cite{Streltsov2010}. We note that by propagating in
imaginary time the MCTDHB equations also allow {one} to determine {the ground 
state} of interacting many-boson systems, {see~\cite{MCHB}}.

\section{Results}
\label{Result}
In this section we discuss {the findings of our study of the dynamics of a BEC in a {one dimensional} ($1\rm{D}$) 
symmetric double well with long-range interaction. Specifically, we examined 
how the properties of the {$1\rm{D}$} BJJ dynamics depend on the range of the inter-atomic interaction}. 
The symmetric double well $V_T(x)$ is constructed by fusing the two harmonic potential 
$V_{\pm}(x)=\frac{1}{2}(x \pm 2)^2$ by a quadratic polynomial $\frac{3}{2}(1-x^2)$ in the region $|x| \le 0.5$. 
We define the width of the double well as the separation between the 
two local minima of $V_T(x)$ given by ${\it l}=4$. The Rabi oscillation{s} in the double well 
sets the time scale for the dynamics, $t_{Rabi}=\frac{2\pi}{\Delta E}=132.498$,
where $\Delta E$ is 
the energy difference between the ground state and the first excited state of a single particle in the trap. 

We start by preparing a BEC of $N$ interacting bosons in 
the ground state of the left well $V_+(x)$ at $t < 0$. At $t \ge 0$, the system is allowed to evolve in time in 
the double well $V_T(x)$. As discussed above, the inter-atomic 
interaction is chosen as {$W(x_j-x_k) = \frac{\lambda_0}{\sqrt{(|x_j-x_k|/D)^{2n}+1}}$} of half-width $D$ with $n=3$. 
The strength of interaction $\lambda_0$ corresponds to 
the mean-field interaction parameter $\Lambda=\lambda_0 (N-1)$. We tune the range $D$ of the interaction for different 
interaction {parameters} $\Lambda$ to study the impact of the long-range tail of the interaction on the BJJ dynamics.
{For all values of $D$ studied here, the peak of the density appears at the centre of the well. This implies that the tail
of the interaction governs the dynamics.}

\begin{figure}[!ht]
\begin{center}
\begin{tabular}{cc}
\includegraphics[width=0.9\linewidth]{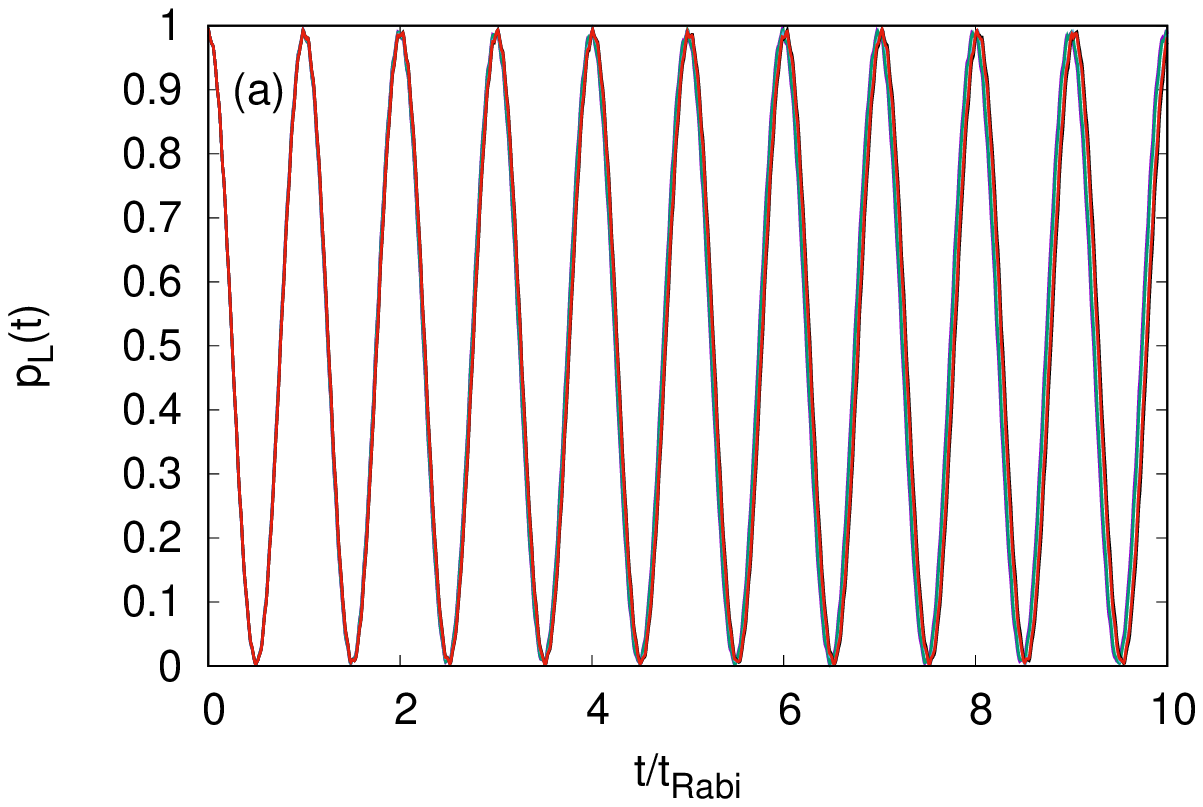} & \\
\includegraphics[width=0.9\linewidth]{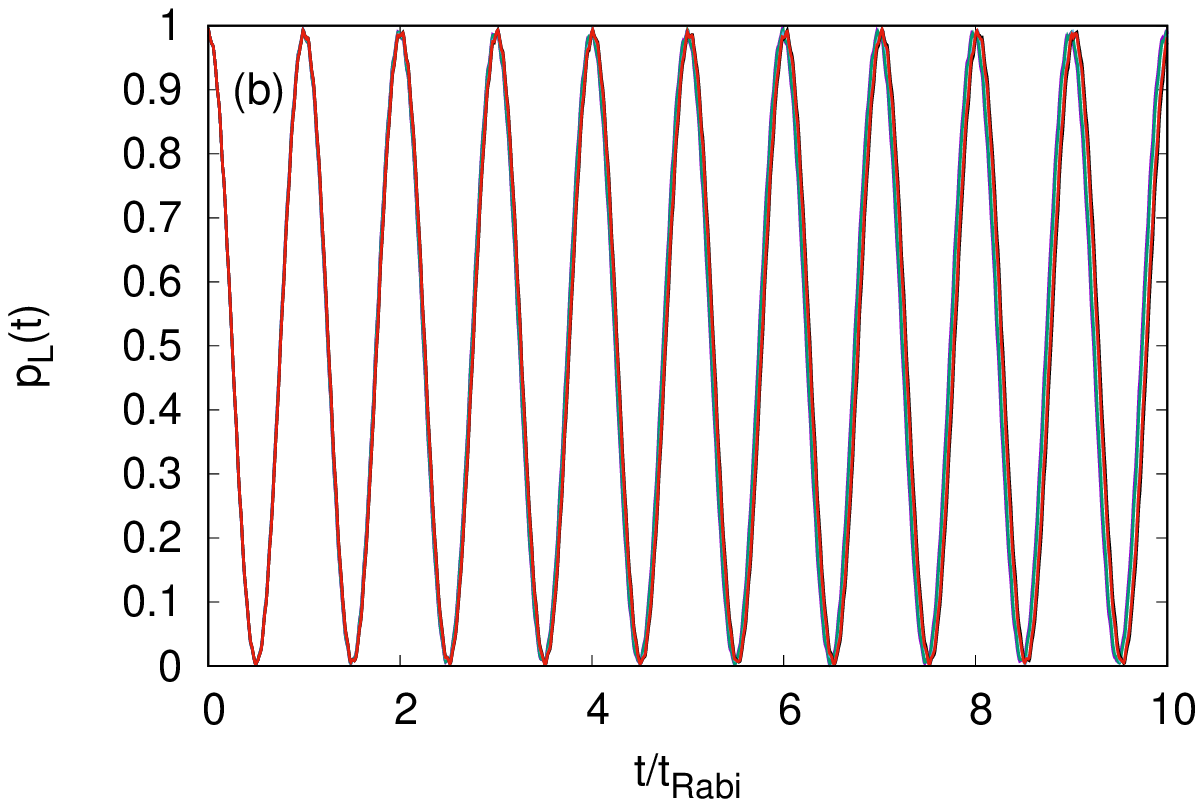} & \\
\end{tabular}
\end{center}
\caption{(color online) {Time evolution} of the survival probability $p_L(t)$ in the left well for various ranges $D$ of 
the interaction for $\Lambda=0.01$. In (a) the mean-field results are
shown while in (b) MCTDHB results with $M=2$ {orbitals} and $N=10000$ {bosons} are shown. The magenta curves correspond to $D=l/8$, 
green curves represent $D=l/4$, blue curves present $D=l/2$, black curves
correspond to $D=3l/4$ and the red curves depict the result for $D=l$; $l$ being the separation between the two 
local minima of the {double-well trap}. See text for further details. 
{The} quantities shown are dimensionless.}
\label{fig.pl-p01}
\end{figure}

We first consider the time evolution of the survival probability in the left well, 
$p_L(t)=\int_{-\infty}^0 {\rm d}x \frac{\rho(x;t)}{N}$, of the BEC in the 1D BJJ. For 
$\Lambda=0.01$, it has been earlier shown {that even for a system with $N=1000$ 
bosons interacting via a contact $\delta$-interaction, the effective interaction is sufficiently weak such that 
the many-body and mean-field results for the time evolution of the density {per particle} and $p_L(t)$ coincide~\cite{Klaiman2016}}. 
Accordingly, here we compute $p_L(t)$ for various $D$ with the mean-field theory and compare them in
Fig.~\ref{fig.pl-p01}(a). We find that for such {a} weak interaction, the system performs full tunnelling 
{oscillations} back and forth between the two wells irrespective of the range of the
interaction{,} and all the curves practically overlap with {each other} though slight deviations appear {at long times}. 
This implies that there is no {pronounced} impact of the long-range tail 
of the interaction on the dynamics {for such {an} interaction {parameter}}. 
The many-body calculations also confirm the same as can be seen from Fig.~\ref{fig.pl-p01} (b) where we present the many-body results for $N=10,000$.
\begin{figure}[!ht]
\begin{center}
\begin{tabular}{cc}
\includegraphics[width=0.45\linewidth]{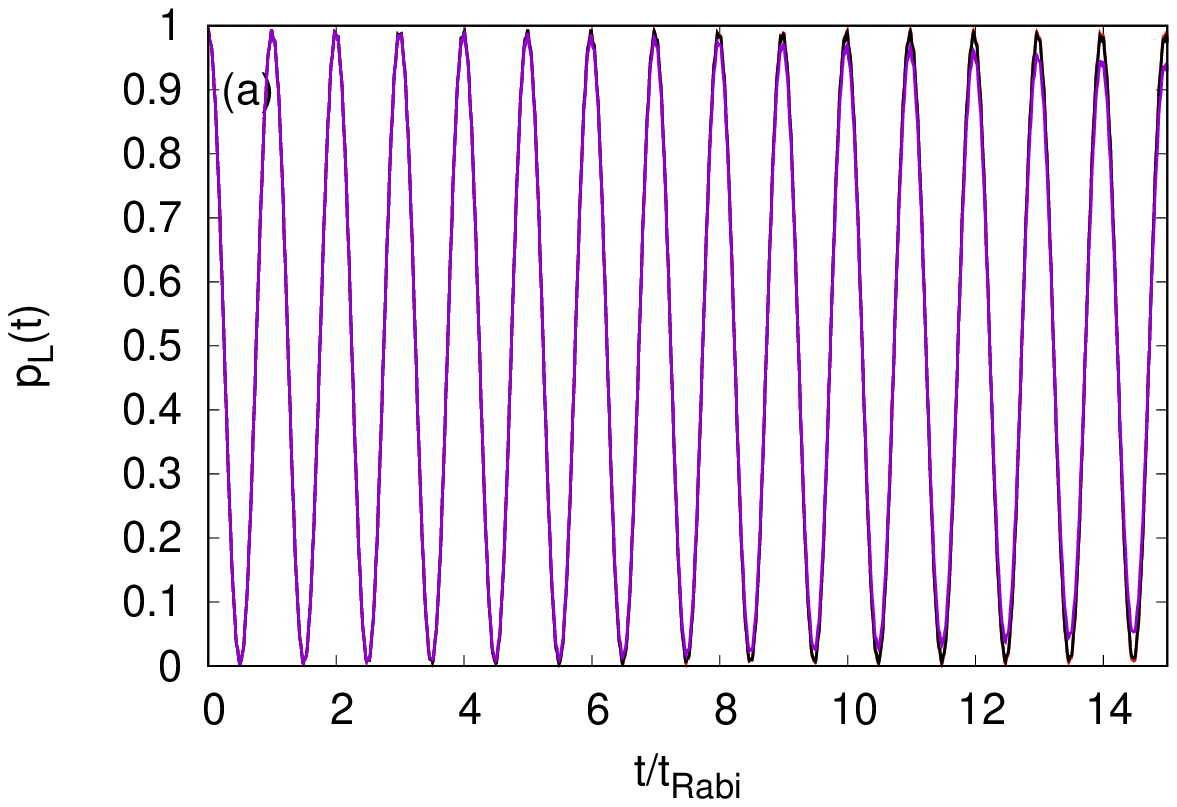} & 
\includegraphics[width=0.45\linewidth]{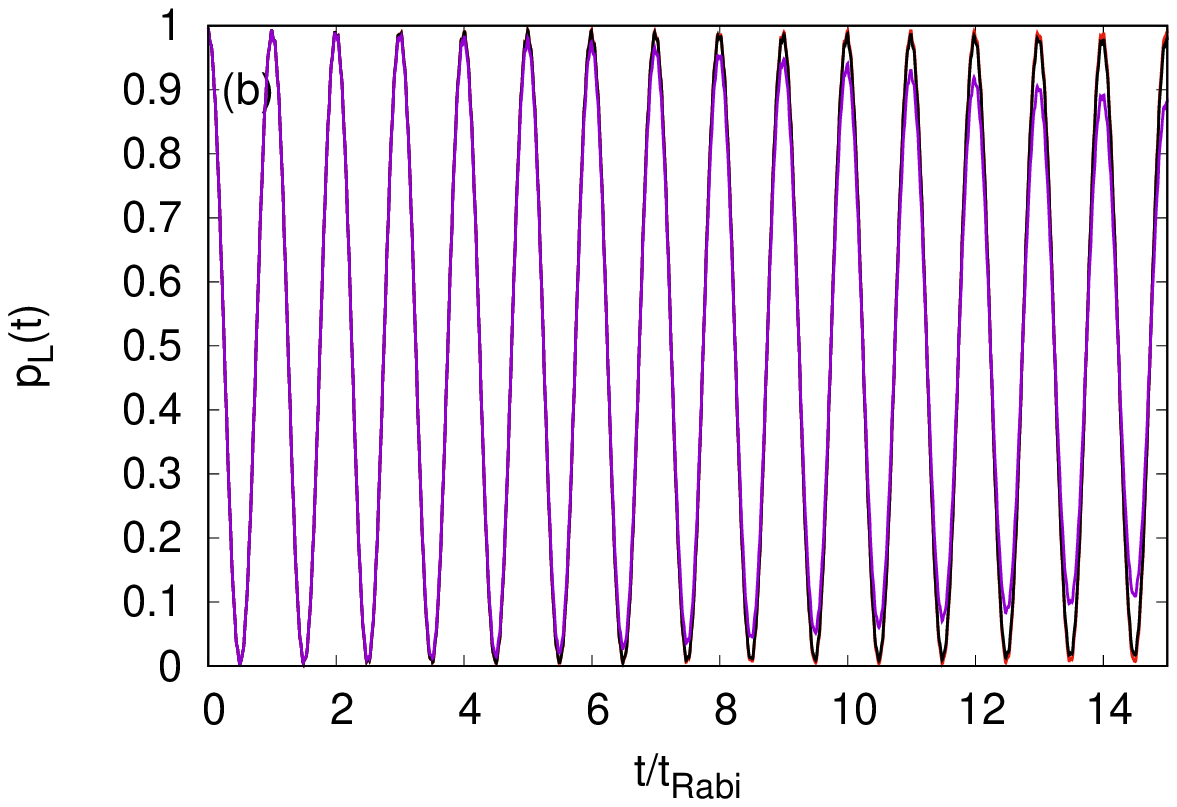}  \\
\includegraphics[width=0.45\linewidth]{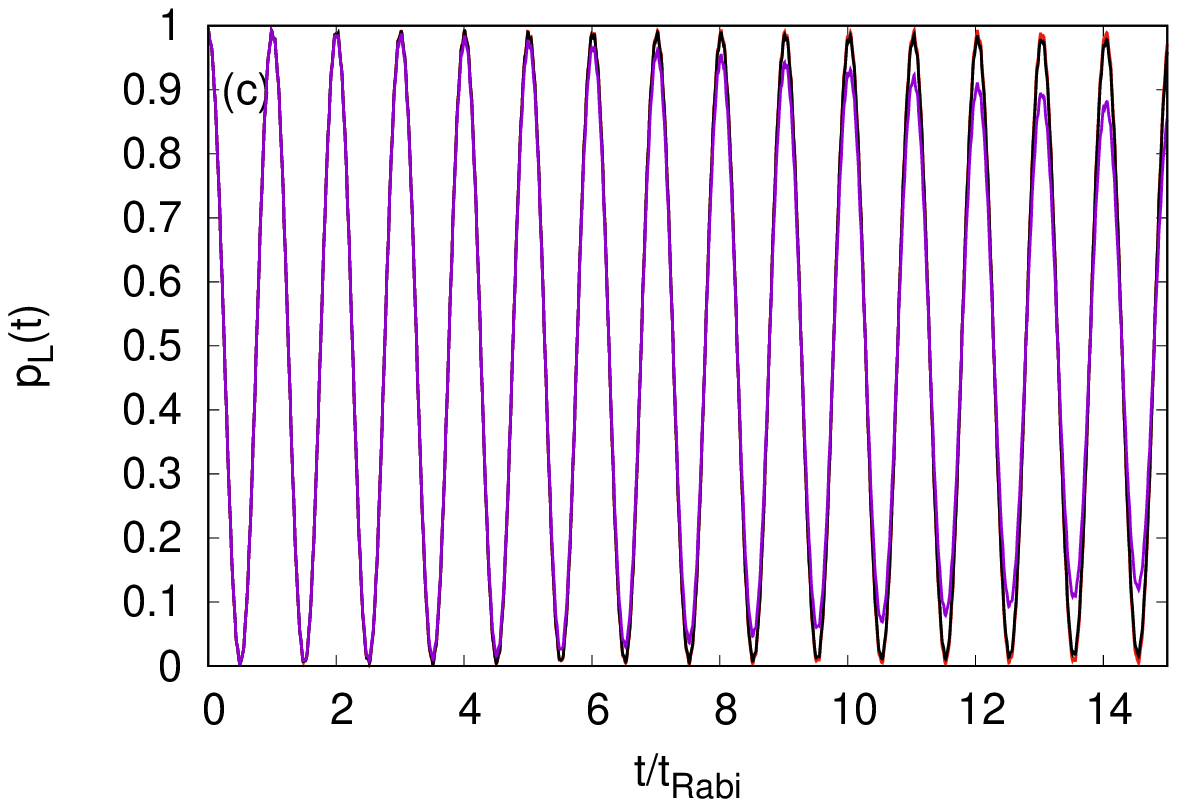} & 
\includegraphics[width=0.45\linewidth]{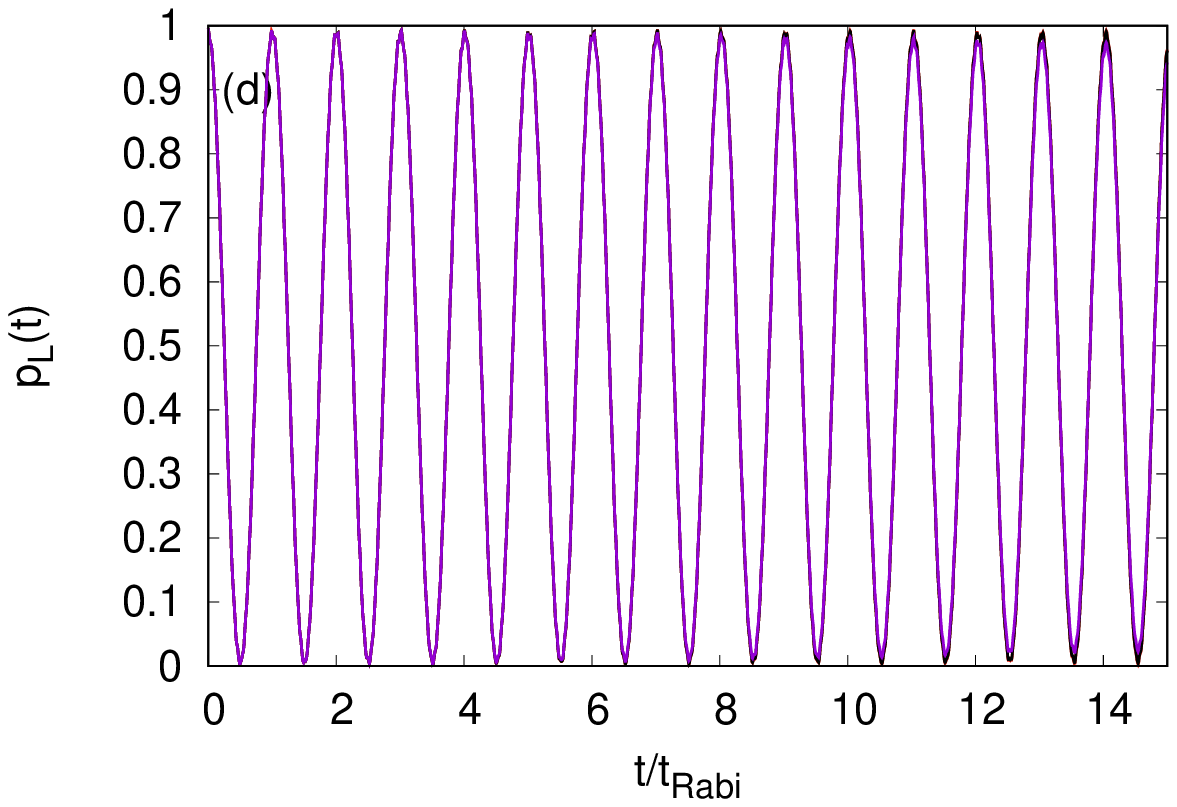} \\
\end{tabular}
\end{center}
\caption{(color online) Time evolution of the survival probability $p_L(t)$ in the left well for various $N$ 
and a fixed range $D$ of the interaction: (a) $D=l/8$, 
(b) $D=l/4$, (c) $D=l/2$ and (d) $D=l$. The interaction parameter is $\Lambda=0.01$. In each panel, the red 
smooth curve corresponds to $N=10000$ {bosons}, black curve to $N=1000$ {bosons} and 
magenta to $N=100$ {bosons}. {The} quantities {shown} are dimensionless.}
\label{fig.pl-p01-D}
\end{figure}
{It is known that in the $N \rightarrow \infty$ limit, keeping $\Lambda$ fixed, the energy per particle and density per particle 
of the system converge to the corresponding mean-field Gross-Pitaevskii (GP) results~\cite{Lieb}{, despite the 
difference between the many-body and GP wave functions~\cite{Klaiman16, Cederbaum2017} } and 
the many-body effects {for these quantities}, if any, would already be erased 
for large $N$. {Similar results hold for the dynamics also~\cite{Erdos2007}.}
Therefore we} have also repeated our many-body 
calculation with smaller number of particles{, viz. $N=100$ and $N=1000$ bosons}. We plot our results in 
Fig.~\ref{fig.pl-p01-D}. We find a competition between the effects of the long-range tail 
and the confining {double-well} trap. For $D \ll {\it l}$, all the
results for different $N$ practically fall on top of each other. However, as $D$ increases we observe a slight damping 
in the oscillation{s} of $p_L(t)$ till $D={\it l}/2$ when the 
damping effect is most prominent. Also, over the same time period, the damping is enhanced for smaller $N$. For larger $D$, 
this effect fades away and finally for $D \geq {\it l}$, it 
completely disappears and the mean-field picture is restored. Further, even though the amplitude of the oscillation{s} 
is damped for smaller $N$, the frequency of the oscillation{s} is 
practically unaffected in all cases and is approximately equal to {the} Rabi {frequency}. This implies that the 
effective interaction is very weak for $\Lambda=0.01$ {and the mean-field limit is
attained already for $N=10,000$ particles as far as $p_L(t)$ is concerned}.    

\begin{figure}[!ht]
\begin{center}
\begin{tabular}{cc}
\includegraphics[width=0.9\linewidth]{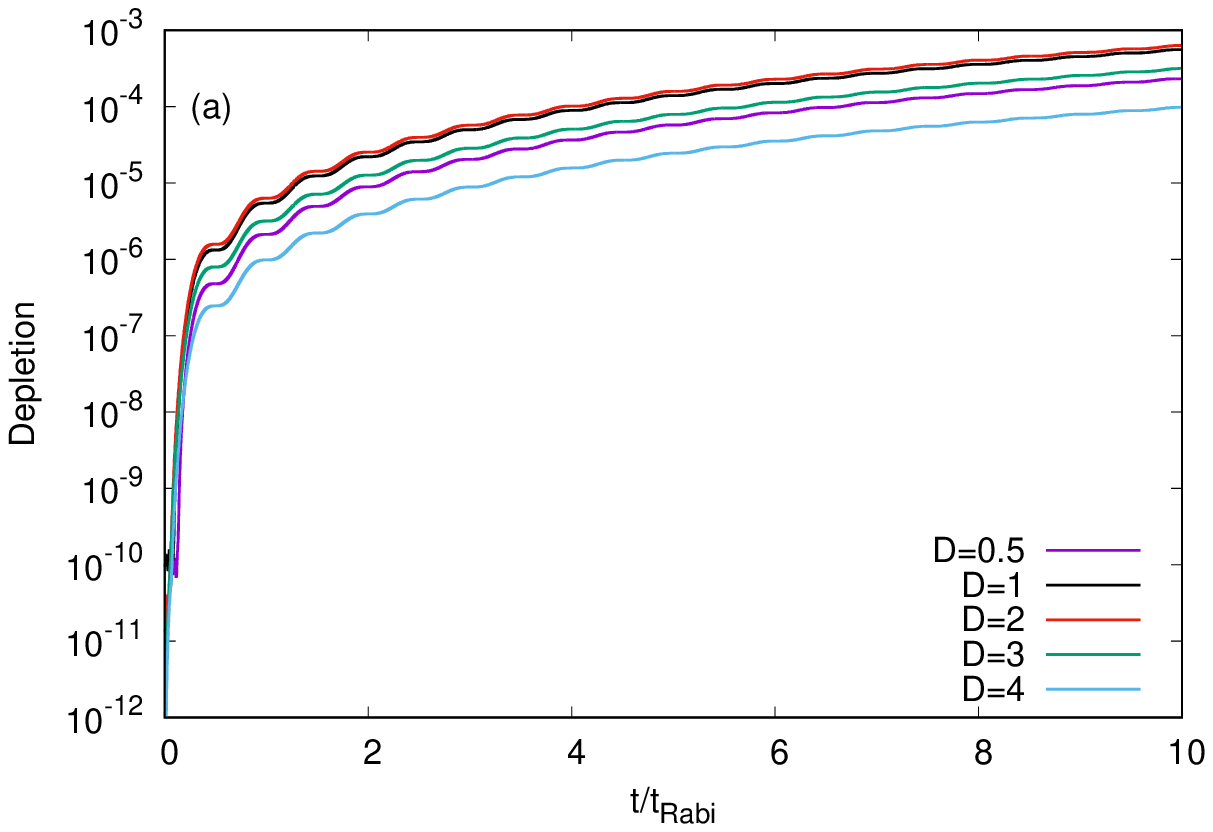} & \\
\includegraphics[width=0.9\linewidth]{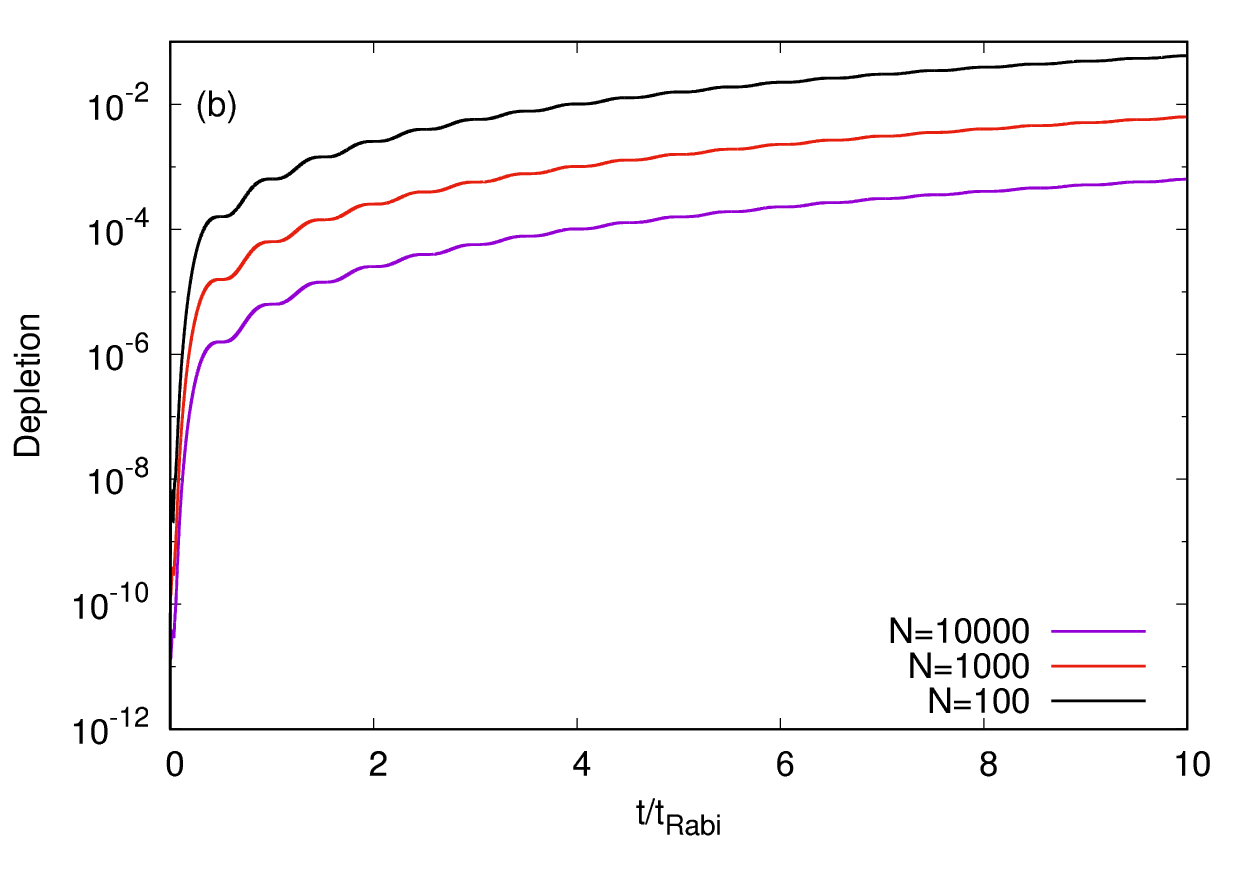} & \\
\end{tabular}
\end{center}
\caption{(color online) (a) The depletion per particle of the condensate as a function of time corresponding to Fig.~\ref{fig.pl-p01}(b). 
(b) The depletion of the condensate {with $N=10,000$ bosons} corresponding to Fig.~\ref{fig.pl-p01-D}(c). 
The respective color code of the curves is 
explained in each panel. {The} quantities shown are dimensionless.}
\label{fig.depletion-p01}
\end{figure}

In order to understand the damping of oscillation{s} of $p_L(t)$ better, we study the depletion $f$ of the system for all three cases. 
In Fig.~\ref{fig.depletion-p01} (a) we plot the depletion
of the BEC with $N=10,000$ {bosons} for various $D$. We find that for such {a small interaction parameter $\Lambda$ the} 
depletion is negligible and the system is essentially {fully} 
condensed irrespective of $D$. Even then, we find
the depletion to increase with $D$ up to $D={\it l}/2$.  This implies that the presence of the {long-range} tail in the 
interacting potential basically enhances the effect of the 
interaction up to $D={\it l}/2$. Moreover, there is a correspondence between the damping of the density oscillation{s and 
the enhancement of the} depletion [see Fig.~\ref{fig.pl-p01-D}]. In 
Fig.~\ref{fig.depletion-p01}(b) we plot {the} depletion of the BEC with different $N$ for $D={\it l}/2$.  
We observe that for {a smaller number of bosons}, the system becomes more depleted and 
therefore {a} more pronounced decay of the oscillation{s of $p_L(t)$ emerges}.

\begin{figure}[!ht]
\begin{center}
\includegraphics[width=0.9\linewidth]{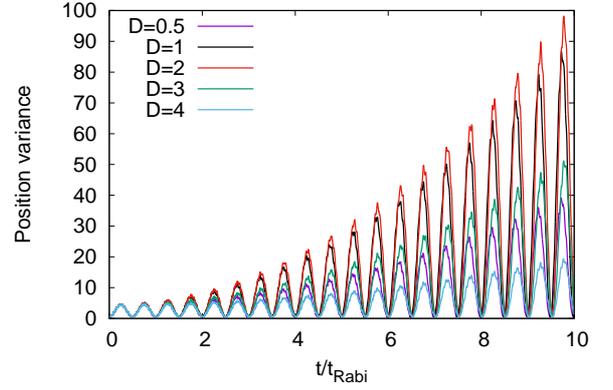}  
\end{center}
\caption{(color online) Time evolution of the position variance per particle $\frac{1}{N} \Delta^2_{\hat X}$ 
for various ranges $D$ of the interaction for $\Lambda=0.01$ and $N=10000$ {bosons}. 
The quantities shown are dimensionless}
\label{fig.variance-p01}
\end{figure}

{The} variance of any operator depends on the actual number of depleted atoms and is a sensitive probe for many-body 
correlation{s}. It can have a large deviation in comparison with the 
mean-field theory even when only one out of a million atoms is out of the condensed {mode}~\cite{Klaiman2015}.
The many-body position variance {per particle $\frac{1}{N} \Delta^2_{\hat X}$ 
of a BJJ and the respective position-momentum uncertainty product have} been found to already deviate from the mean-field 
result even for such {a} weak {$(\Lambda=0.01)$} interaction strength~\cite{Klaiman2016}.  
So, it can serve as a better probe to investigate how the many-body correlations are affected in the system due to 
the presence of the long-range tail in the interaction. 
Moreover, in Fig.~\ref{fig.depletion-p01}(a) we observe that the curves {of the depletion} for different $D$ are shifted 
vertically from each other suggesting that {for a given $N$}, 
the total number of atoms residing in higher {natural} orbitals depends on the range of the interaction $D$. 
Therefore, {the} variance of {an} operator is expected 
to exhibit strong dependence on $D$. 
Thus, next we consider the variance {per particle} of the many-body position operator $\hat X=\sum_{j=1}^N \hat x_j$ 
of the system which {can be expressed as follows}~\cite{Klaiman2016, Klaiman2015}
\beqn\label{dis}
\frac{1}{N}\Delta_{\hat X}^2 &=& \frac{1}{N} 
\left(\langle\Psi|\hat X^2|\Psi\rangle - \langle\Psi|\hat X|\Psi\rangle^2\right) \nonumber \\
&=& \int d\r \frac{\rho(\r)}{N}x^2  - N \left[\int d\r \frac{\rho(\r)}{N}x\right]^2  \nonumber \\
&+& \sum_{jpkq} \frac{\rho_{jpkq}}{N(N-1)} \cdot (N-1) \int d\r_2 \, 
\phi^{\ast{NO}}_j(\r_1) \phi^{\ast{NO}}_p(\r_2) \, \nonumber \\
& &\times \, x_1x_2 \, \phi^{NO}_k(\r_1) \phi^{NO}_q(\r_2). \
\eeqn
In Fig.~\ref{fig.variance-p01}, we plot the variance of the many-body position operator of a BEC of 
$N=10,000$ bosons for different $D$. For all cases, {$\frac{1}{N} \Delta^2_{\hat X}$} 
is found to grow in an oscillatory manner, at least for a few Rabi cycle{s}. However, the growth 
rate depends on $D$ in an intricate manner. {The variance} is found to grow faster 
with $D$ until $D={\it l}/2$ and then for further increase in $D$, the growth of {$\frac{1}{N} \Delta^2_{\hat X}$} becomes slower. 
Also, even as the maximal values of {$\frac{1}{N} \Delta^2_{\hat X}$} grows with time, its minima values always
return to approximately zero. Since the system is essentially {fully} condensed for $\Lambda=0.01$, 
in each oscillation the system is {practically} localized once in each of the two wells 
leading to near zero values for the minima. The slight deviation from the zero {values can be attributed} to the finite depth of the double well. 
On the other hand, as the number of depleted atoms out
of the condensate varies with $D$, the maximal values of {the variance} over the same period of time and consequently 
the growth rate vary accordingly (with $D$).  
Moreover, as for {the survival probability} $p_L(t)$, here also the frequency of oscillations is 
{nearly} unaffected and is equal to twice the Rabi frequency for the double well.  

\begin{figure}[!ht]
\begin{center}
\begin{tabular}{cc}
\includegraphics[width=0.9\linewidth]{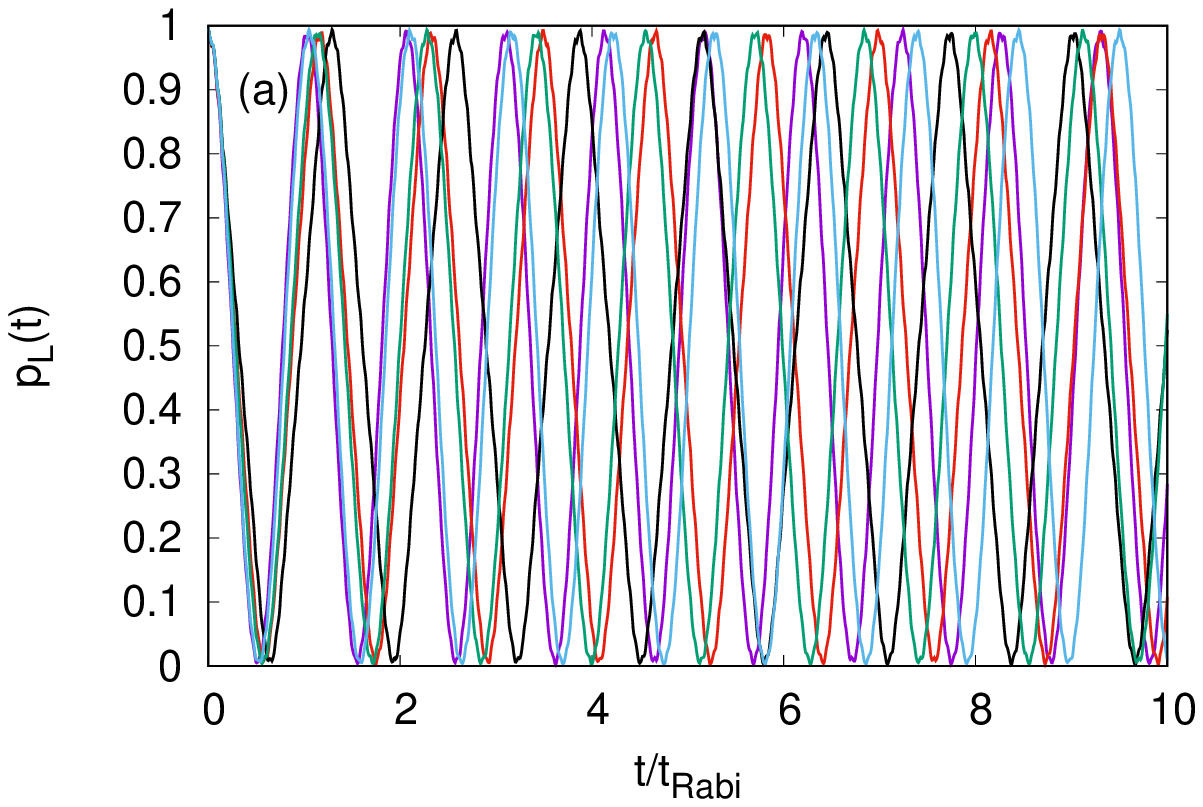} & \\ 
\includegraphics[width=0.9\linewidth]{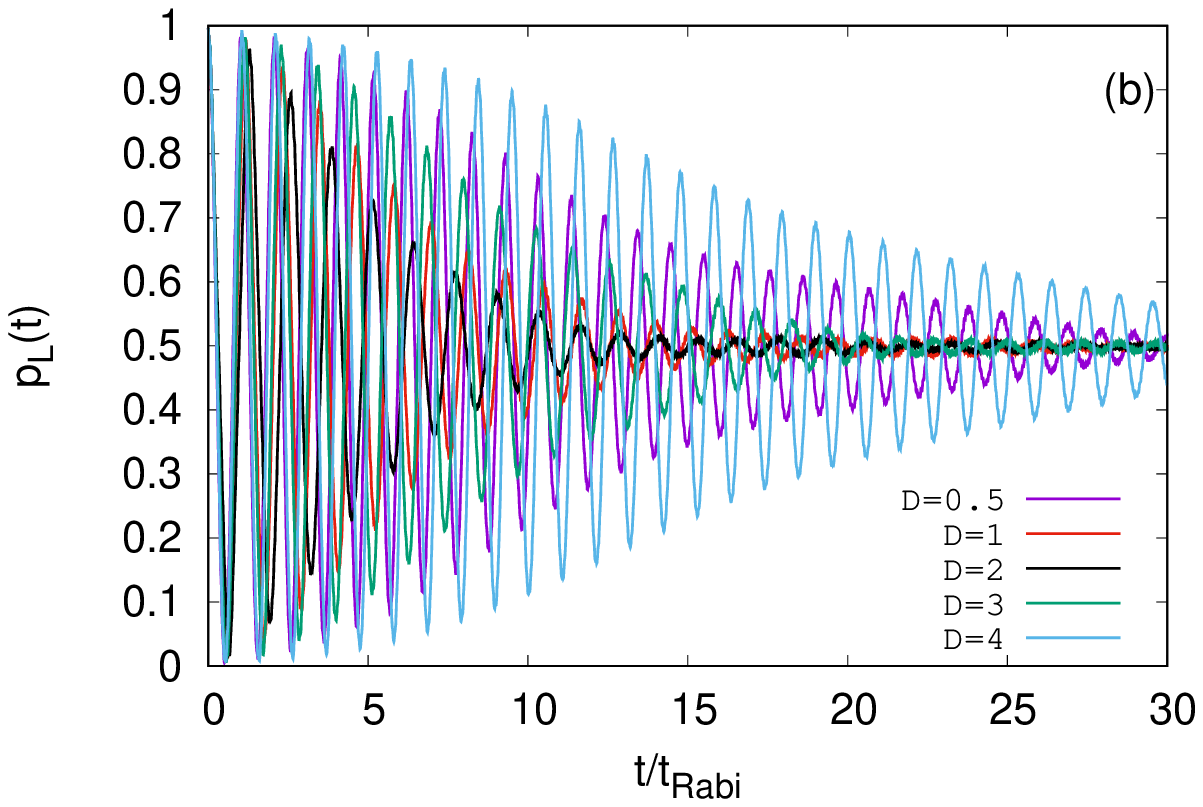}  &\\
\end{tabular}
\end{center}
\caption{(color online) The time evolution of the survival probability $p_L(t)$ in the left well for various ranges $D$ of 
the interaction for $\Lambda=0.1$. In (a) the 
mean-field results are shown while in (b) MCTDHB results with $M=2$ {orbitals} and $N=1000$ {bosons are} shown. 
The color code of the curves is explained in panel (b). The quantities shown are dimensionless.}
\label{fig.pl-p1}
\end{figure}

To explore the damping in the density oscillations further, we next consider {a ten times stronger interaction} 
$\Lambda=0.1$. Here also, the mean-field result for $p_L(t)$ shows 
[Fig.~\ref{fig.pl-p1}(a)] that the
density of the system tunnels back and forth between the two wells irrespective of the range $D$ of the interaction. 
However, the frequency of oscillations is found to {visibly}
depend on $D$, {compare to Fig.~\ref{fig.pl-p01}}. The frequency is found to {somewhat} increase with $D$ up to 
$D={\it l}/2$ and then {it} slowly decreases for larger $D$. 
So, within the mean-field theory, we already observe some effect of the long-range tail of the interaction. 

As the mean-field theory cannot describe the collapse and revival of the density oscillation{s}, we also calculate 
$p_L(t)$ in the left well for a
system of $N=1000$ bosons by {the} MCTDHB method with $M=2$ orbitals. We plot the many-body results in Fig.~\ref{fig.pl-p1}(b) 
for various $D$. We find that, although the density of the system
tunnels back and forth between the two well{s}, the amplitude of the oscillation{s} decreases with time and eventually it collapses 
for all $D \leq {\it l}$. However, the collapse time
$t_{collapse}$ is found to depend on the range $D$ of the interaction. It is observed that the collapse occurs earlier as $D$ 
increases from a small value up to $D={\it l}/2$ and then
$t_{collapse}$ increases for further increase in $D$. In {\cite{Sakmann2014}}, it has been shown for contact interaction that 
$t_{collapse}$ decreases for larger $\Lambda$ with $N$ 
fixed. This implies that {up to $D=l/2$,} the presence of the long range tail of the interaction enhances the effect of 
{the} interaction compared to the contact interaction with {a} similar strength. 

\begin{figure}[!ht]
\begin{center}
\includegraphics[width=0.9\linewidth]{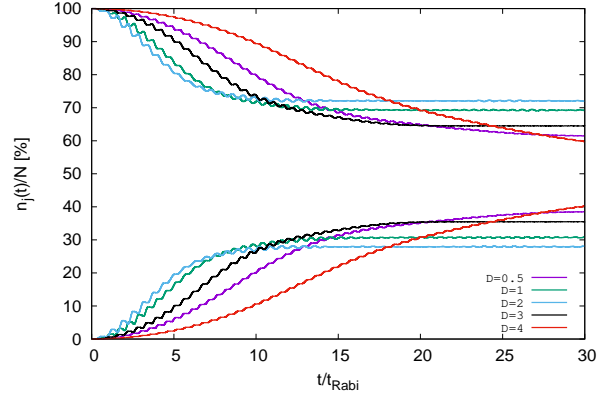}  
\end{center}
\caption{(color online) The fragmentation of the condensate with time for different ranges $D$ for the interaction parameter 
$\Lambda=0.1$ and $N=1000$ {bosons. See text for further details. The} quantities shown are dimensionless.}
\label{fig.frag-p1}
\end{figure}

To further explore the connection between the depletion or fragmentation and the collapse of the density oscillation{s} and 
its dependence on $D$, next we study the corresponding natural
occupations. We stress that fragmentation of a condensate is a pure many-body phenomena and cannot be described by the mean-field 
GP theory~\cite{Klaiman2006}. 
In Fig.~\ref{fig.frag-p1}, we plot the natural occupations for different $D$. Since we started with a practically fully condensed state, 
initially only one natural orbital is 
occupied with occupancy {$\frac{n_1}{N}\approx1$} and the corresponding fragmentation $f = \frac{n_2}{N} \approx 0$. 
However, with time as the condensate starts to tunnel through the 
barrier, the second natural orbital becomes occupied and 
the condensate becomes fragmented. The fragmentation $f$ reaches a plateau $f=f_{col}$ around $t_{collapse}$. We found that along with 
$t_{collapse}$, $f_{col}$ also depends on $D$. 
$f_{col}$ first decreases as $D$ is increased starting from a small value till $D={\it l}/2$, and then for larger $D$ it increases. 

\begin{figure}[!ht]
\begin{center}
\includegraphics[width=0.9\linewidth]{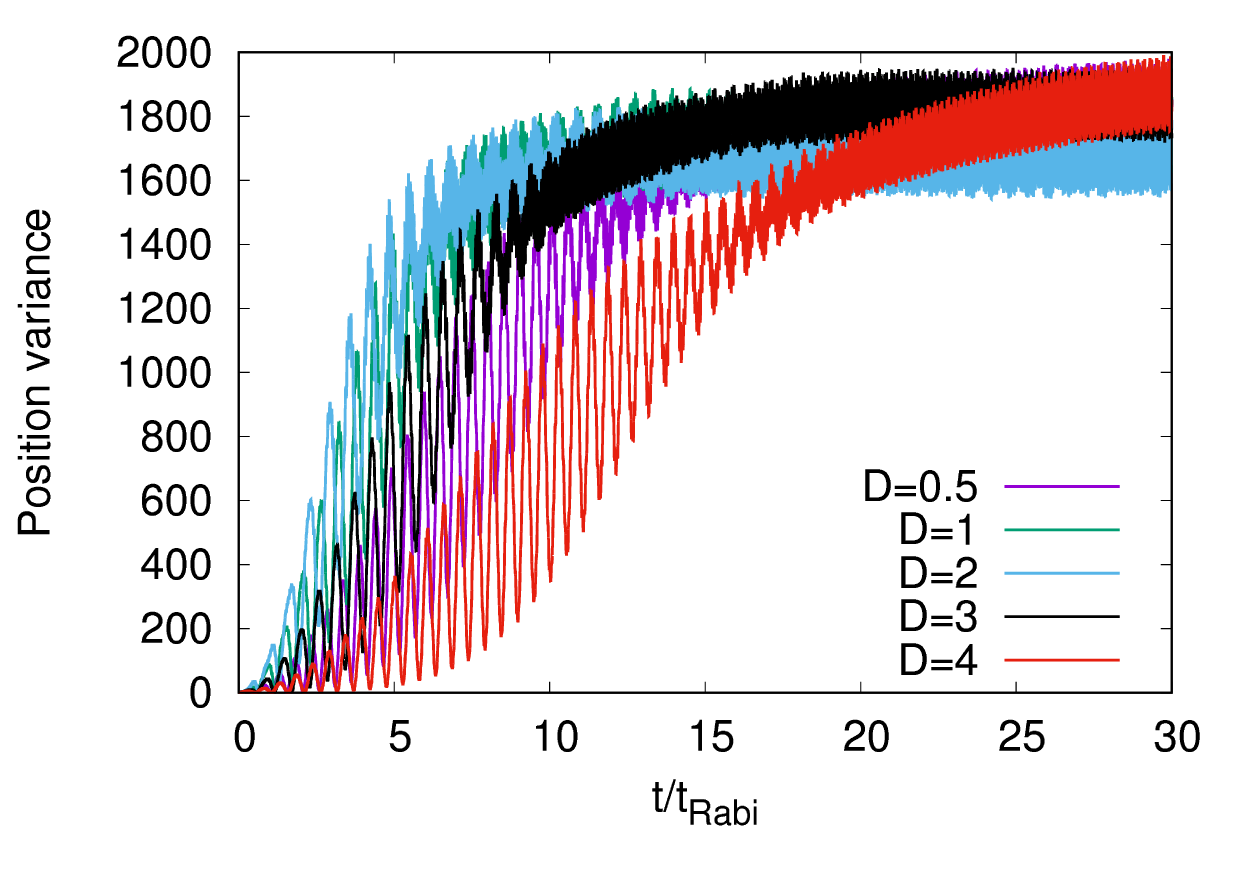}  
\end{center}
\caption{(color online) The time evolution of the position variance per particle $\frac{1}{N} \Delta^2_{\hat X}$ for various ranges 
$D$ of the interaction $\Lambda=0.1$ and $N=1000$ {bosons. For further details see text.}  
The quantities shown are dimensionless.} 
\label{fig.variance-n1k-p1}
\end{figure}

Now that the condensate is fragmented, {the} behavior of {the position variance} is {even} more interesting. 
In Fig.~\ref{fig.variance-n1k-p1}, we plot 
{$\frac{1}{N} \Delta^2_{\hat X}$} as a function of time for various $D$. Here also we found {the variance} to increase 
with time in an oscillatory manner for all $D$. 
Moreover, the pace of growth of the variance is again {seen} 
to vary with $D$ in a similar manner as in the case of $\Lambda=0.01$. However, there is a fundamental difference in this case. 
Now, not only the maxima values but also the minima
values of {$\frac{1}{N} \Delta^2_{\hat X}$} increase. This is {a consequence} of the growing {degree of} fragmentation of the BEC. 
Finally at $t \sim t_{collapse}$, as the density oscillations collapse and {the}
fragmentation $f$ reaches a plateau at $f_{col}$, {the} variance also exhibits an equilibration{-like} effect and 
oscillates about a constant mean value. 
As $f_{col}$ shifts vertically with $D$,
the actual number of depleted atoms outside {the} condensate {mode} also varies {with} $D$. Naturally, the mean 
value about which {$\frac{1}{N} \Delta^2_{\hat X}$} 
oscillates also moves up and down similarly to $f_{col}$, {see Fig.~\ref{fig.variance-n1k-p1}}.
\begin{figure}[!ht]
\begin{center}
\includegraphics[width=0.9\linewidth]{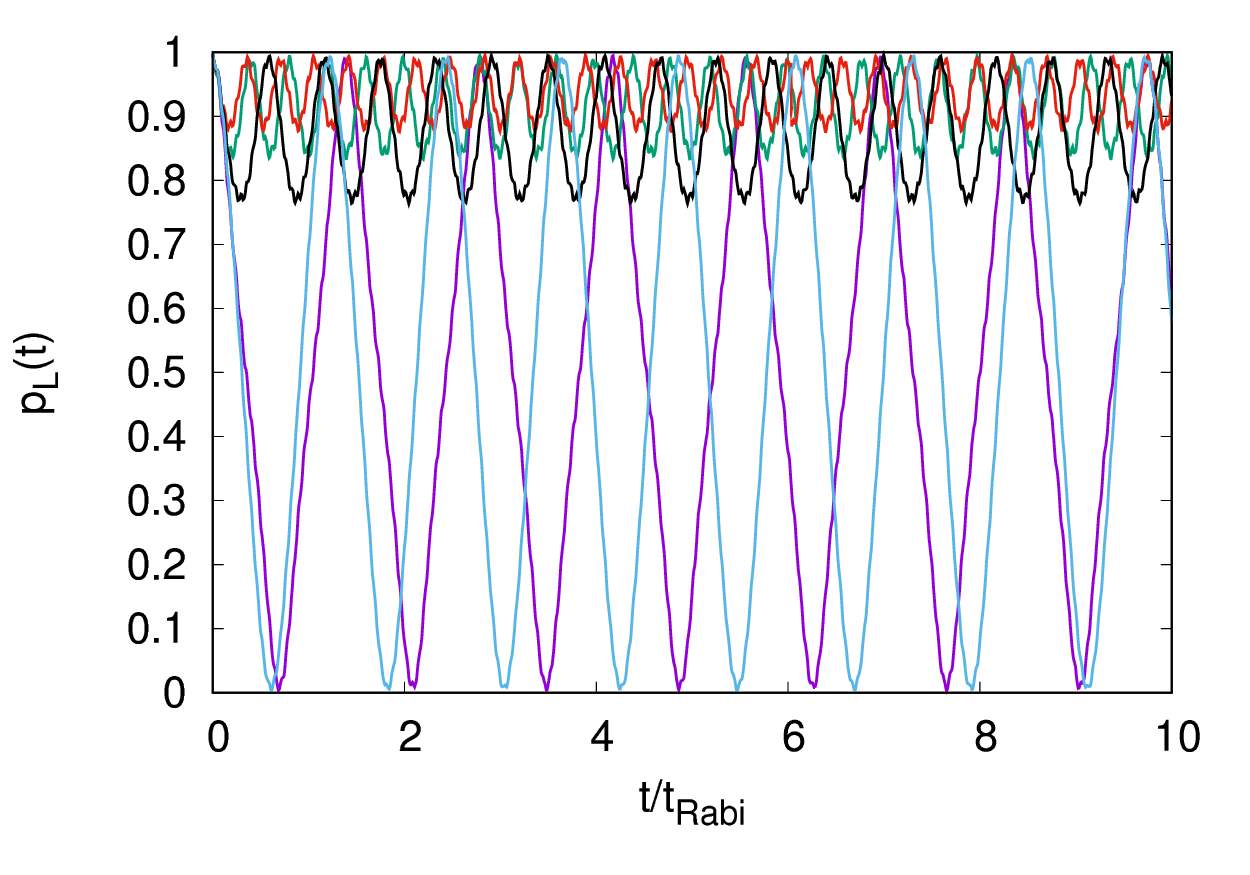} 
\end{center}
\caption{(color online) The time evolution of the survival probability $p_L(t)$ in the left well for various ranges $D$ of the interaction. 
{Only the mean-field results for the interaction parameter $\Lambda=0.2$ are shown.}
The magenta smooth curve corresponds to $D=l/8$, green curve represents $D=l/4$, red curve presents $D=l/2$, 
black curve corresponds to $D=3l/4$ and the blue curve depicts
the result for $D=l$. For further details see text. The quantities shown are dimensionless.
}
\label{fig.pl-p2}
\end{figure}

{Last,} to examine how the frequency of the density oscillation{s} {and the resulting survival probability itself are}
affected by the range of the interaction more prominently, 
we increase $\Lambda$ further. {Here we remind that} {self trapping} is an
important feature in the BJJ dynamics with {$\delta$}{-}interaction~\cite{Milburn, Shenoy} 
when the interaction is stronger than a critical value. 
In this case, a {BEC} initially localised in one well tends to remain {self trapped} in that
well for more than a few Rabi cycles. {So, next, we double the strength of interaction to $\Lambda=0.2$.} 
In Fig.~\ref{fig.pl-p2} we plot the mean-field result
for $p_L(t)$. We find that now, not only the frequency but also the amplitude of the density oscillation{s} is 
{strongly} affected due to the presence of a tail part in the interaction. 
For small $D$, similarly to zero-range contact $\delta$-interaction, the system performs full oscillations back and 
forth between the two well{s}. However, for larger $D$, the system exhibits
partial {self trapping} with only up to about $20\%$ tunneling to the other well. Finally, for $D \approx l$, 
the complete oscillations of $p_L(t)$ is restored. This partial 
{self trapping}, {as far as we know}, is a novel phenomenon found in presence of a long-range tail in the interaction potential
{already at the mean-field level. We leave it as a future task to investigate the many-body facets of this self-trapping phenomenon}. 

\section{Conclusions}
\label{conclusion}

In short, here we have studied how the well known BJJ dynamics is modified in presence of a long-range tail in 
the interaction potential. We choose a model two-body interaction 
{$W(x_i-x_j)$} with a tunable range $D$. Such interactions are relevant in view of recent experiments 
with dressed Rydberg atoms and ultra-cold
atoms with {a} strong dipole {moment}. The qualitative physics discussed here is expected to remain the same for other 
repulsive interactions with similar strength{s} and range{s}.
We studied the impact of the presence of a tail part in the interaction potential {at} both the mean-field {and 
many-body levels}. For the many-body calculation{s}, we used the MCTDHB
method with $M=2$ orbitals. We have also numerically checked that for our present study, MCTDHB computations with 
$M=2$ orbital{s are} {accurate} and the convergence of our findings with
respect to $M$ {time-adaptive orbitals} is discussed in {the} Appendix.

We examined the well-known aspects of BJJ dynamics, viz., density oscillation{s and their} collapse, {self trapping}, 
depletion and fragmentation {as well as} the {recently examined} position variance. 
We again stress that{,} while the
density oscillation{s} and self trapping already have some explanations {at} the mean-field {level}, the collapse 
and revival of the density oscillations and fragmentation can only be 
described by a many-body theory. Moreover, {the} variance is a {sensitive} many-body quantity which depends on the actual 
number of depleted atoms and hence, {can} deviate from the mean-field result
even in the $N\rightarrow \infty$ limit. The impact of the range of the interaction on the dynamics is manifested through 
several features{: the} frequency of {the} density oscillation{s;}
depletion; collapse time $t_{collapse}$; fragmentation plateau $f_{col}$; maximal and minimal values of the position 
variance in each cycle of {the} oscillation{s} and the growth pace of the 
variance{. These} are
key to our study.

Our {work} revealed that for {a} very weak interaction, the system is essentially {fully} condensed and 
the density oscillation{s are} practically
unaffected by the presence of the tail in the interaction potential. However, as the strength of the interaction 
increases, its impact gradually becomes 
prominent. We found an intricate competition between { the {double-well} trap and the range of the interaction}. 
Initially, as the range of the interaction is
increased from a small value, it is tantamount to pushing the atoms further {away} from each other. {Naturally, increasing} 
the range of the interacting 
potential effectively enhances the repulsive
interaction. However, as the range of the interaction becomes comparable with the width of the {double well} trap, 
the contribution of the confining effect of the trap enhances and the system
behaves as if all the atoms are in a constant {interaction potential}. This, in effect, diminishes the impact of 
the range of interaction on the dynamics of the system. 
Therefore, in this study, we have
{successfully shown that the use of a long-range interaction enriches the physics of the BJJ dynamics}. Since BJJ is a paradigmatic
device for understanding coherent quantum phenomena with applications in precision measurements~\cite{Orzel} and sensing~\cite{Hall}, 
our present study is of fundamental importance. 

\section*{Acknowledgement}

This paper is dedicated to Professor Hans-Dieter Meyer, a dear colleague and friend, on the occasion of his 70th birthday. 
This research was supported by the Israel Science Foundation (Grant No. 600/15). We are grateful to Shachar Klaiman and Alexej Streltsov for discussions. 
Computation time on the High Performance Computing system Hive of the Faculty of Natural Sciences at the University of Haifa and on the Cray XC40 
system Hazelhen at the High Performance Computing Center Stuttgart (HLRS) is gratefully acknowledged. SKH gratefully acknowledges 
the continuous hospitality of the Lewiner Institute for Theoretical Physics (LITP) at the Department of Physics, Technion - Israel Institute of Technology.

\appendix
\section{MCTDHB computations and their accuracy}\label{I}

As already discussed in Sec.~\ref{Theory}, the ansatz in MCTDHB is taken as the superposition of $M$ orbitals 
which are determined by a time-dependent variational 
principle. 
In this connection, we would like to mention that in the limit $M \to \infty$ the set of permanents $\{\left|\vec{n};t\right>\}$
spans the complete $N$-boson Hilbert space and thus expansion (\ref{MCTDHB_Psi}) is exact. 
On the other hand, for $M=1$ we get back the {well-known GP} equation.
However, in actual numerical calculations we have, of course, to limit the size of the Hilbert 
space exploited. By allowing also the permanents to be time-dependent we can use much shorter expansions than if only
the expansion coefficients are taken to be time-dependent, thus leading to a significant computational advantage.

In our numerical {calculations}, {the many-body Hamiltonian is represented by 128 exponential discrete-variable-representation
grid points (using a Fast Fourier transformation routine) in a box of size[-10,10). We obtain the initial state for the time 
propagation - the many-body ground state of the BEC in the left well, 
by propagating the MCTDHB equations of motion in imaginary time~\cite{MCHB, Lode2012}. For our numerical computations
we use the numerical implementation in the software Package~\cite{Streltsov1,Streltsov2}.} 
We keep on {repeating the computation with} increasing $M$ until convergence is reached 
{and thereby obtain the numerically-exact results}.

As a concrete example, convergence with increasing $M$ of the time-dependent many-particle position variance 
{ per particle $\frac{1}{N} \Delta^2_{\hat X}$ of $1\rm{D}$} BJJ is {discussed here}.
As discussed in the text, the many-particle position variance 
is more sensitive to many-body effects compared to the oscillations in {survival probability 
$p_L(t)$ and the} fragmentation $f$. Hence, convergence of {$\frac{1}{N} \Delta^2_{\hat X}$} with increasing $M$ should also 
imply the convergence of $p_L(t)$ and $f$ with 
respect to $M$; {see also ~\cite{Klaiman2016, Klaiman2015}}. 
\begin{figure}[!ht]
\begin{center}
\begin{tabular}{cc}
\includegraphics[width=0.45\linewidth]{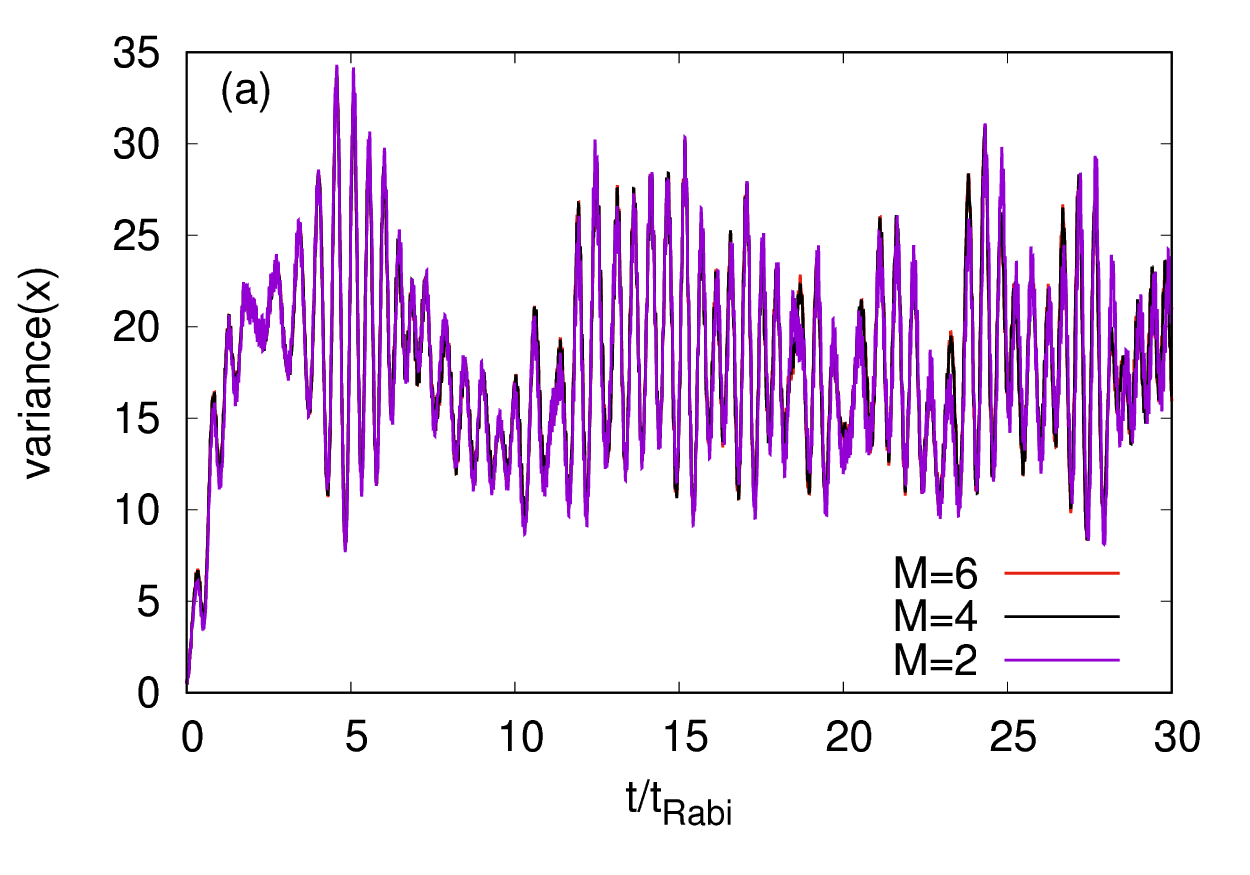} & 
\includegraphics[width=0.45\linewidth]{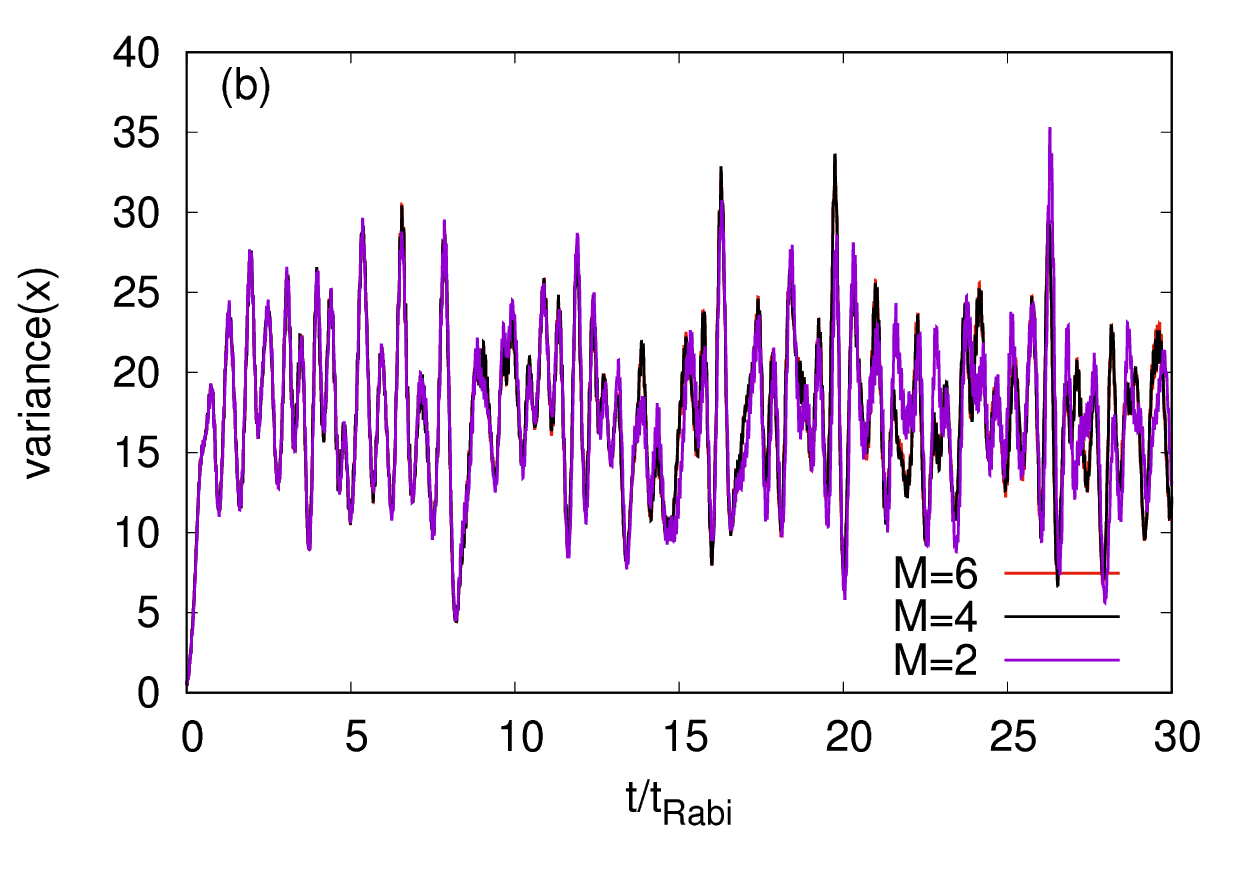}  \\
\includegraphics[width=0.45\linewidth]{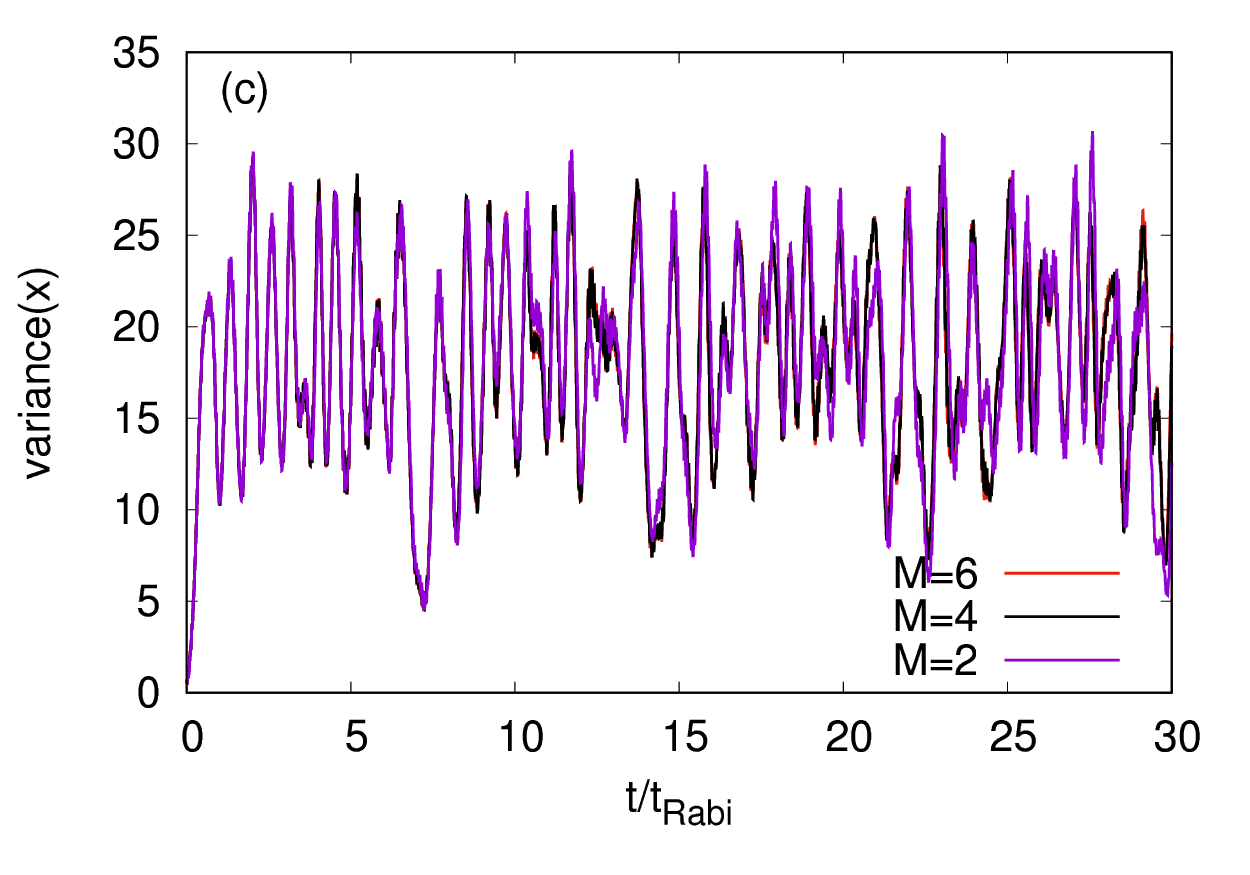} & 
\includegraphics[width=0.45\linewidth]{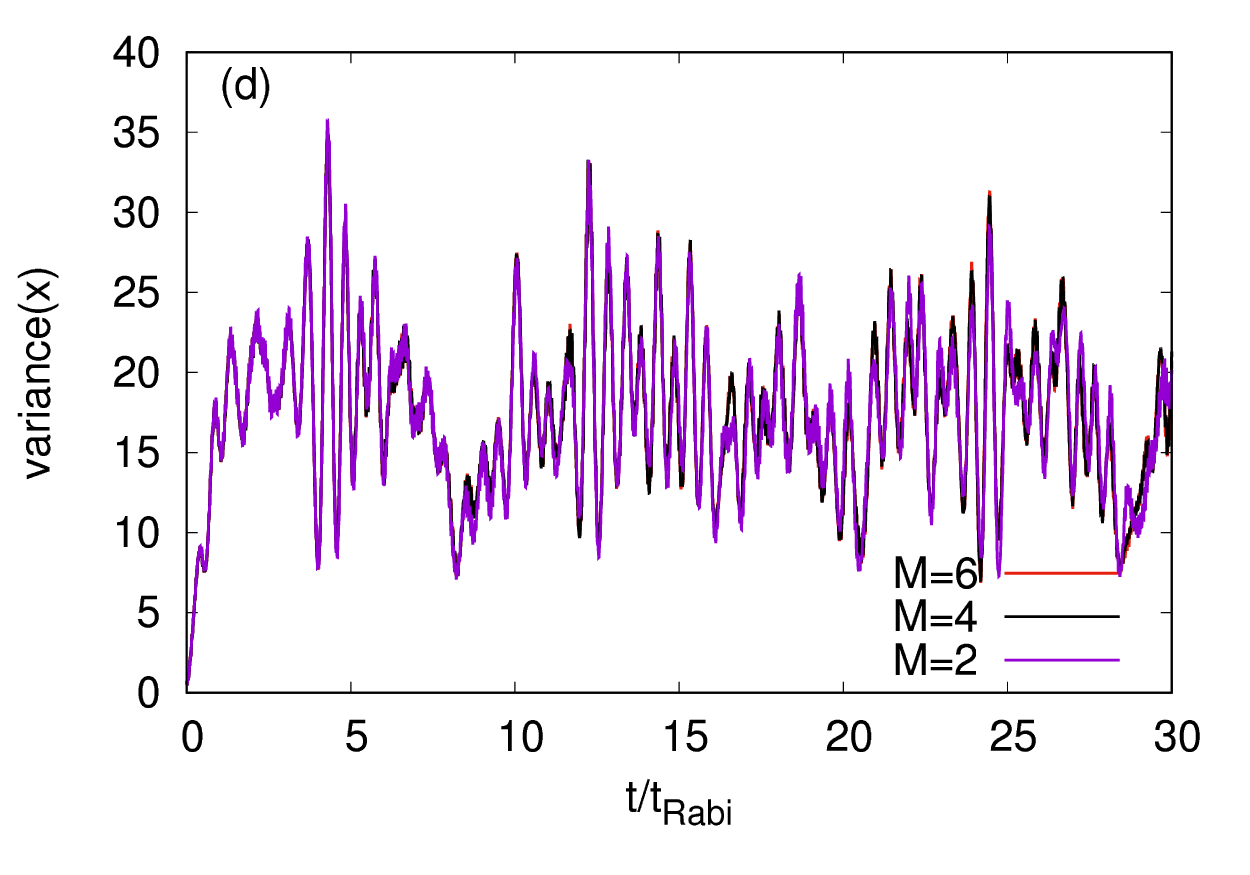} \\
\end{tabular}
\includegraphics[width=0.5\linewidth]{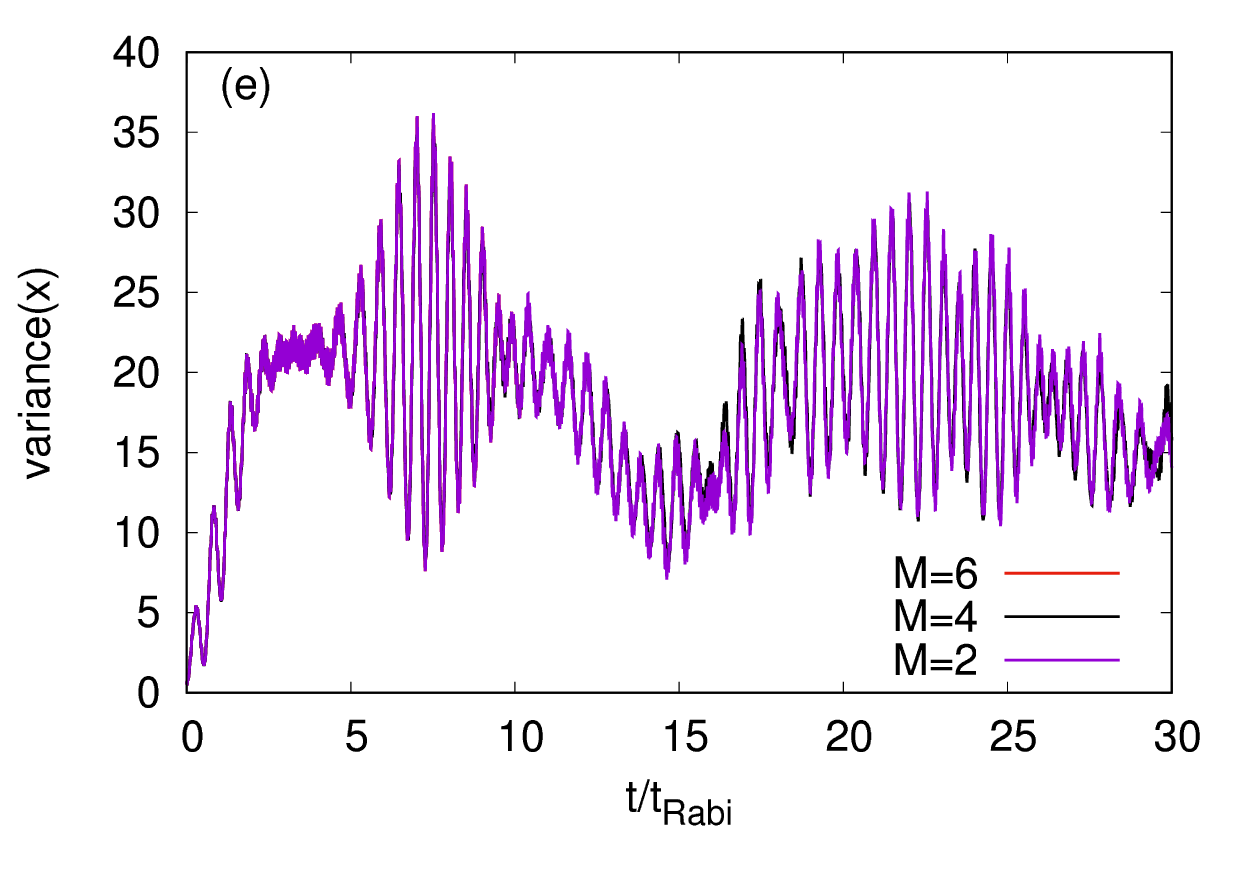} \\
\end{center}
\caption{(color online) Convergence of the time-dependent many-body position variance {per particle}
of a BEC in the one dimensional BJJ with increasing $M$ is shown for {for 
$N=10$ bosons and interaction parameter $\Lambda=0.1$.} (a) $D=l/8$; (b) $D=l/4$; 
(c) $D=l/2$; (d) $D=3l/4$ and (e) $D=l$. {The results with $M=2$ accurately describe the variance while 
the results with $M=4$ and $M=6$ orbitals} lie on top of each other.
See text for further details.
The quantities shown are dimensionless.}
\label{fig.conv-p1}
\end{figure}

{In Fig.~\ref{fig.conv-p1}, we have plotted the MCTDHB results for $\frac{1}{N} \Delta^2_{\hat X}$ of $1\rm{D}$ 
BJJ computed with $M=$2, 4 and 6 orbitals for $\Lambda=0.1$ and $N=10$. We} see that, 
similarly to Fig.~\ref{fig.variance-n1k-p1}, there is an equilibration{-like} effect in  
{$\frac{1}{N} \Delta^2_{\hat X}$} though now the oscillations of {$\frac{1}{N} \Delta^2_{\hat X}$} about {a fixed mean value} 
are strongly aperiodic due to 
{the rather} small system size. {Moreover, it can be seen that all the curves practically lie top on each other for all $D$. 
In fact, the MCTDHB results for $M=4$ and $M=6$ almost completely overlap each other for all $D$.
Also, as can be seen from Fig.~\ref{fig.conv-p1}(a) and (d) the overlap between the 
curves is near perfect for very small $D$ as well as for $D \approx l$. 
Thus, our MCTDHB computations with $M=2$ orbitals are already converged and aptly describe the physical behavior of the system, 
at least for the $\Lambda$ and period of the dynamics considered here}.

Since increasing $N$, keeping the mean-field parameter $\Lambda$ fixed, amounts to {a} weaker interaction strength 
{$\lambda_0$}, convergence of our results 
for $N=1000$ is actually expected to be better than Fig.~\ref{fig.conv-p1}. Obviously, it should improve further for {the} 
smaller {interaction parameter} $\Lambda=0.01$. 
Indeed, form Fig.~\ref{fig.pl-p01} we see that the survival probability $p_L(t)$ {(and hence, also the density per particle)}
converges already at the {GP} level for $\Lambda=0.01$ {and $N=10,000$}.


\end{document}